\documentclass[twocolumn,prl,superscriptaddress,nofootinbib]{revtex4-1}

\usepackage{graphicx}
\usepackage{amsmath,bm}
\usepackage[usenames,dvipsnames]{xcolor}
\usepackage[normalem]{ulem}
\usepackage{url}
\usepackage{multirow}

\usepackage[colorlinks  = true,
            linkcolor   = NavyBlue,
            urlcolor    = NavyBlue,
            citecolor   = NavyBlue,
            anchorcolor = NavyBlue]{hyperref}

\def \aap  {A\&A}

\def \apj  {ApJ}

\def \apjl  {ApJL}

\def \araa  {ARA\&A}
\def \jcap  {JCAP}

\def \ssr {SSRv}
\def \mnras {MNRAS}

\def \jgr {JGR}

\def \prl {PRL}
\def \planss {Planetary and Space Science}

\newcommand\thbn{\vartheta_{\rm Bn}}
\renewcommand{\deg}{^{\circ}}
\newcommand{\vsh}{v_{\rm sh}}
\newcommand{\esh}{E_{\rm sh}}
\newcommand{\Em}{E_{\rm max}}
\newcommand{\omci}{\omega_c^{-1}}

\begin{document}

\title{Fast particle acceleration in 3D hybrid simulations of quasi-perpendicular shocks}

\author{Luca Orusa}
\affiliation{Department of Physics, University of Torino, via P. Giuria, 1, 10125 Torino, Italy}
\affiliation{Istituto Nazionale di Fisica Nucleare, via P. Giuria, 1, 10125 Torino, Italy}
\author{Damiano Caprioli}
\affiliation{Department of Astronomy \& Astrophysics, University of Chicago, Chicago, IL 60637, USA}
\affiliation{Enrico Fermi Institute, The University of Chicago, Chicago, IL 60637, USA}

\begin{abstract}
Understanding the conditions conducive to particle acceleration at collisionless, non-relativistic shocks is important for the origin of cosmic rays.
We use hybrid (kinetic ions---fluid electrons) kinetic simulations to investigate particle acceleration and magnetic field amplification at non-relativistic, weakly magnetized, quasi-perpendicular shocks. So far, no self-consistent kinetic simulation has reported non-thermal tails at quasi-perpendicular shocks. Unlike 2D simulations, 3D runs show that protons develop a non-thermal tail spontaneously (i.e., from the thermal bath and without pre-existing magnetic turbulence). They are rapidly accelerated via shock drift acceleration up to a maximum energy determined by their escape upstream. We discuss the implications of our results for the phenomenology of heliospheric shocks, supernova remnants and radio supernovae.

\end{abstract}

\maketitle


\section{Introduction}

Understanding the conditions conducive to particle acceleration at collisionless, non-relativistic shocks is important for the origin of cosmic rays and for understanding the phenomenology of heliospheric shocks \cite[e.g.,][]{mason+99,turner+21}, novae \cite[e.g.,][]{metzger+16,diesing+23}, supernova remnants \cite[e.g.,][]{morlino+12,caprioli12} (SNRs), winds and lobes of active galaxies \cite[e.g.,][]{dermer+09,matthews+19}, and galaxy clusters \cite{brunetti+14}.

Energization at shocks proceeds mainly via diffusive shock acceleration (DSA) \cite{krymskii77,axford+78, bell78a, blandford+78}:
particles are scattered back and forth across the discontinuity, gaining energy via a series of first-order Fermi cycles, due to head-on collisions with magnetic field irregularities. 
The maximum attainable energy is determined by the rate of scattering and hence on the level of magnetic perturbations \cite{lagage+83a,blasi+07,bell+13}.

To self-consistently take into account the non-linear interplay between particle injection/acceleration and self-generated magnetic turbulence at non-relativistic shocks, numerical kinetic simulations are necessary. 
Full particle-in-cells (PIC) simulations have captured ion and electron DSA at quasi-parallel shocks, i.e., when the angle between the background field ${\bf B}_0$ and the shock normal is $\thbn\lesssim 45\deg$ \cite{park+15, crumley+19, shalaby+22}.
Simulations of (quasi-)perpendicular shocks ($\thbn\approx 90\deg$) have been performed in 1D \cite[e.g.,][]{shimada+00,kumar+21,xu+20}, 2D \cite[e.g.,][]{amano+09a,bohdan+21,kato+10,matsumoto+15}, and 3D \cite[e.g.,][]{matsumoto+17}, but evidence of DSA has been elusive.
Also at relativistic shocks DSA is more efficient for quasi-parallel configurations \cite{sironi+09, sironi+11}, unless the shock  magnetization is sufficiently low, which makes DSA possible, though rather slow \cite{sironi+13}.

Full-PIC simulations can cover only a limited range of time/space scales.
Hybrid simulations retain ions as kinetic particles and consider electrons as a neutralizing fluid \cite[e.g.,][]{winske+96,lipatov02}, which effectively removes the displacement current from Maxwell's equation (Darwin approximation).
Hybrid codes do not need to resolve the small time/length scales of the electrons, which are usually dynamically negligible, and are thus better suited than full PIC codes to simulate the long-term evolution of shocks.
2D runs have shown that thermal ions can be \emph{spontaneously} injected into DSA at quasi-parallel shocks
\cite{kucharek+91,giacalone+93,giacalone+97,burgess+05,caprioli+14b,caprioli+14c, caprioli+15,caprioli+17,caprioli+18,haggerty+20,caprioli+20}, but injection of ions at oblique and quasi-perpendicular shocks has been more problematic.
Test-particle and Monte Carlo calculations with prescribed strong scattering seem conducive to ion injection \cite{baring+95,giacalone+00,giacalone03}, but no self-consistent kinetic simulation has reported DSA tails.
Nevertheless, hybrid simulations \emph{augmented} with upstream magnetic fluctuations with long-wavelength and large amplitudes \cite{giacalone05} suggest that quasi-perpendicular shocks may be efficient ion accelerators. 
In general, when  magnetic turbulence is seeded by hand in the pre-shock medium \cite{giacalone+00, guo+13}, or when energetic seeds are added \citep{caprioli+18}, DSA may occur for arbitrary inclinations, though with efficiency and spectra that are not universal but depend critically on the \emph{ad-hoc} prescriptions for the pre-existing seeds and turbulence. 

It has been recognized that systems with reduced dimensionality artificially suppress particle diffusion across field lines \cite{jokipii+93,jones+98,giacalone+00}, which likely explains why 1D/2D simulations fail to produce non-thermal tails.
The question remains of whether, and under which conditions, 3D cross-field diffusion of supra-thermal particles can lead to \emph{spontaneous} injection into DSA.

In this Letter we use hybrid simulations to explore quasi-perpendicular shocks from the magnetized to the weakly-magnetized regime, finding that also for non-relativistic shocks ion acceleration is naturally unlocked if the ordered magnetic field is sufficiently low and the full 3D dynamics is retained. 

\section{Simulation setup}
\begin{figure}[t]
\begin{center}
    \includegraphics[width=0.485\textwidth]{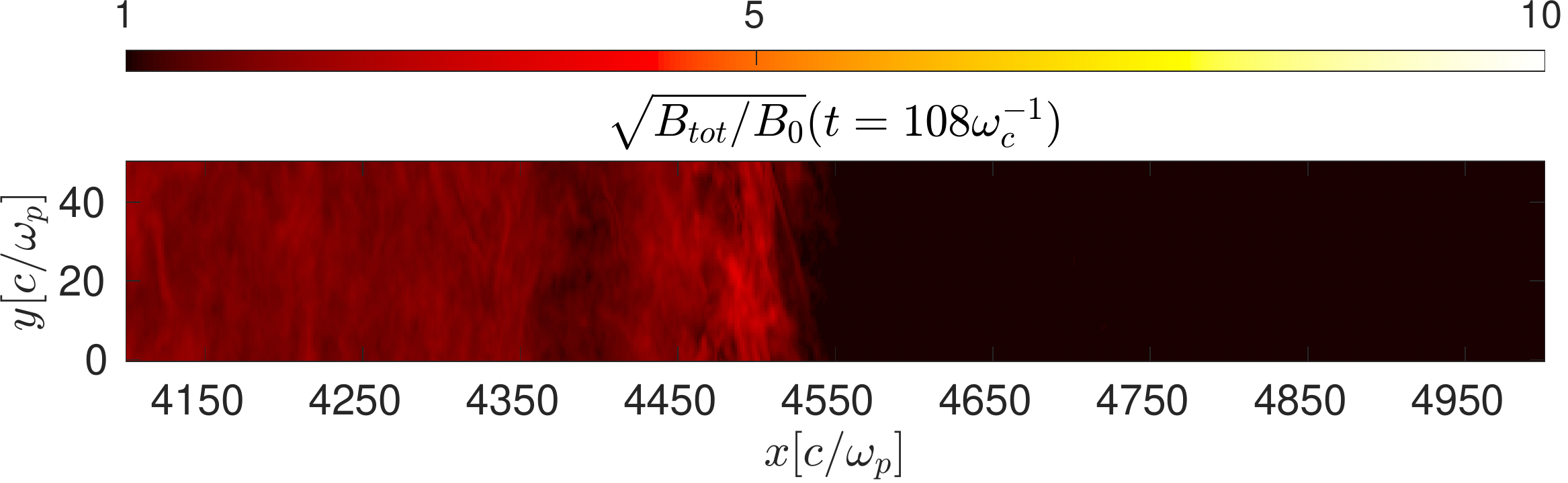}
    \\
    \includegraphics[width=0.48\textwidth]{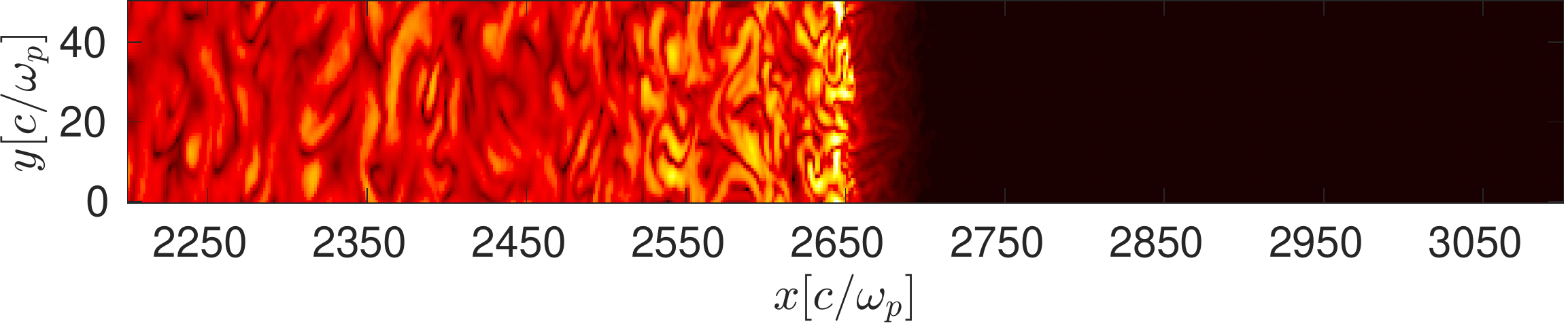}
    \\
    \includegraphics[width=0.48\textwidth]{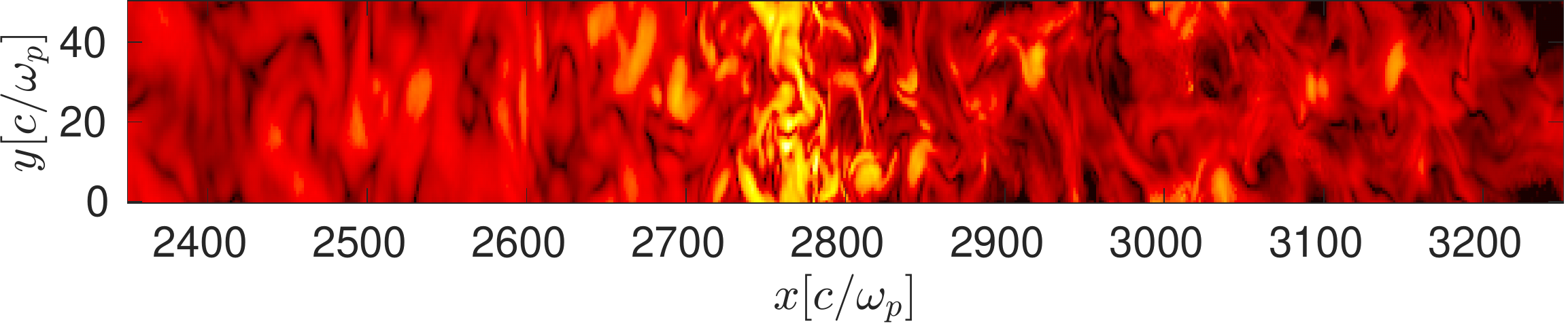}
    \includegraphics[width=0.49\textwidth]{B_evolution.pdf}
    \end{center}
    \caption{From the top: Total magnetic field for the out-of-plane (2D$z$), in-plane (2D$y$), and 3D setups. Bottom panel: evolution of $B_{\rm tot}$ in 3D. In all cases $M$=100 and $t=108 \omega_c^{-1}$.} 
    \label{Fig:B_field}
\end{figure}

\begin{table}[b]
\caption{Left: Parameters for the runs in the Letter; inclination is fixed to $\thbn=80\deg$. 
Right: Corresponding acceleration efficiency, $\varepsilon$, and energy spectral index, $\alpha$.}
\label{Table}
\begin{tabular}{ c c c c | c c c }
 \hline \hline
 Run & $M$ & $x$[$d_i$] &  $\Delta t$[$\omega_c^{-1}$] &  $\varepsilon$ ($> 10\esh$) &  $\alpha$ \\
\hline 
A   &   20    &    5000   &   7.5 $\times 10^{-3}$ &   4\%  &  3.7 \\
B   &   30    &    5000   &   4 $\times 10^{-3}$ &   8\%  &  3.1\\
C   &   40    &    5000   &   1.5 $\times 10^{-3}$ &   11\% &  2.9\\
D   &   50    &    5000   &   1.5 $\times 10^{-3}$ &   17\% &  2.7\\
E   &   100   &    10000   &   7.5 $\times 10^{-4}$ &   30\%  &  1.5\\
 \hline \hline
\end{tabular}
\end{table}

Simulations are performed with the {\tt dHybridR} code \cite{haggerty+19a}, here used in the non-relativistic regime \cite{gargate+07}.
We send a supersonic flow with speed  $\vsh$ against a reflecting wall (left boundary), which produces a shock moving right into a quasi-perpendicular ${\bf B}_0$ field with $\thbn=80\deg$.

Lengths are measured in units of the ion skin depth $d_i\equiv c/\omega_p$, where $c$ is the light speed and $\omega_p\equiv \sqrt{4\pi n e^2/m}$ is the ion plasma frequency, with $m,e$ and $n$ the ion mass, charge and number density, respectively. 
Time is measured in inverse cyclotron times $\omega_c^{-1}\equiv mc/(eB_0)$. 
Velocities are normalized to the Alfv\'en speed $v_A\equiv B/ \sqrt{4\pi m n}$ and energies to $\esh\equiv m \vsh^2/2$.
Simulations include the three spatial components of the particle momentum and of the electromagnetic fields.
Ions are initialized with thermal velocity $v_{\rm th} = v_A$ and electrons are an adiabatic fluid initially in thermal equilibrium.
We define the sonic Mach number as $M_{\rm s} \equiv \vsh / c_{\rm s}$, with $c_{\rm s}$ the speed of sound and the Alfv\'enic Mach number as $M_A\equiv\vsh/v_A$; 
throughout the Letter we indicate the shock strength simply with $M = M_A \simeq M_s$.

We performed simulations with different $M$, longitudinal sizes, and time steps, as reported on the left side of Table~\ref{Table}; 
the transverse sizes are fixed to $50 d_i$ in all the cases.
We use two and a half cells per $d_i$ in each direction and 8(4) ion particles per cell (ppc) in 3D(2D).
We checked our results against convergence in particle statistics, box size, and spatial/temporal resolution:  e.g., increasing the transverse sizes to 100 $d_i$ and ppc to 16 returns energy spectra indistinguishable from those presented here. 
For discussions on the convergence in box size we refer to the appendix of \cite{caprioli+14a,caprioli+14b}, for the hybrid equation of state to \cite{haggerty+20,caprioli+18}, and for linear and strongly non-linear problems simulated in 2D and 3D to \cite{haggerty+19a}.

\section{Results}\label{Results}
We start by comparing the dynamics of quasi-perpendicular shocks in 1D, 2D with $\mathbf{B}$ either in plane (2D$y$) or out of plane (2D$z$), and 3D. 

\begin{figure}[t]
\begin{center}
    \includegraphics[width=0.48\textwidth]{"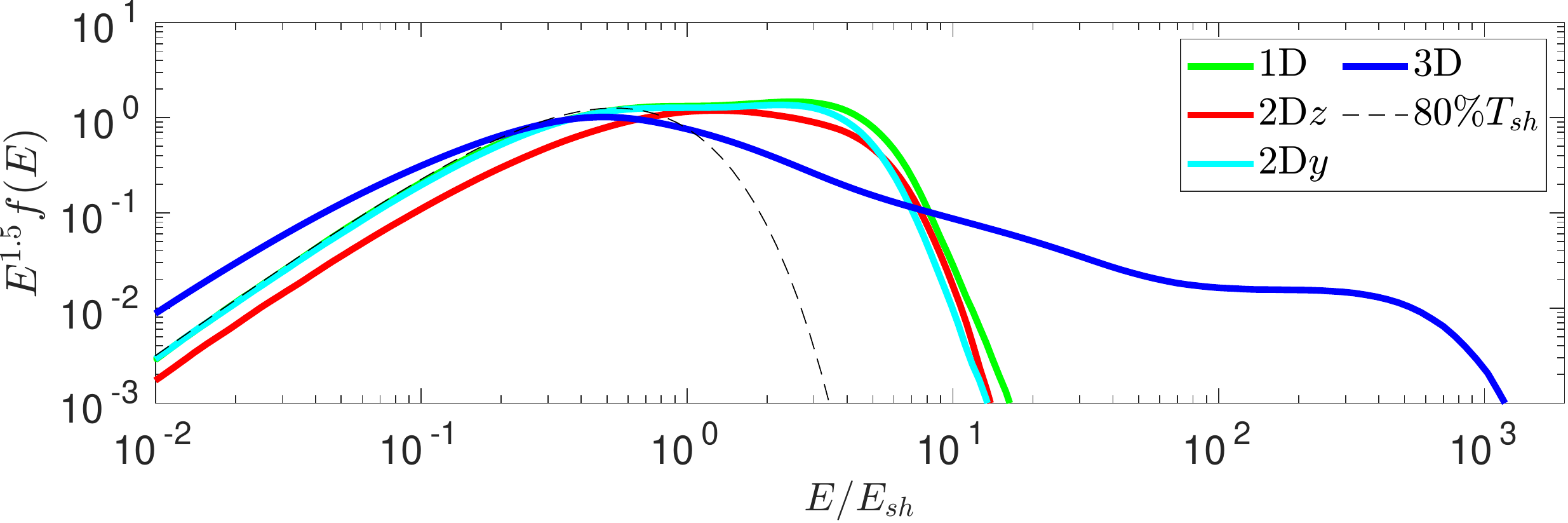"}
    \includegraphics[width=0.48\textwidth]{"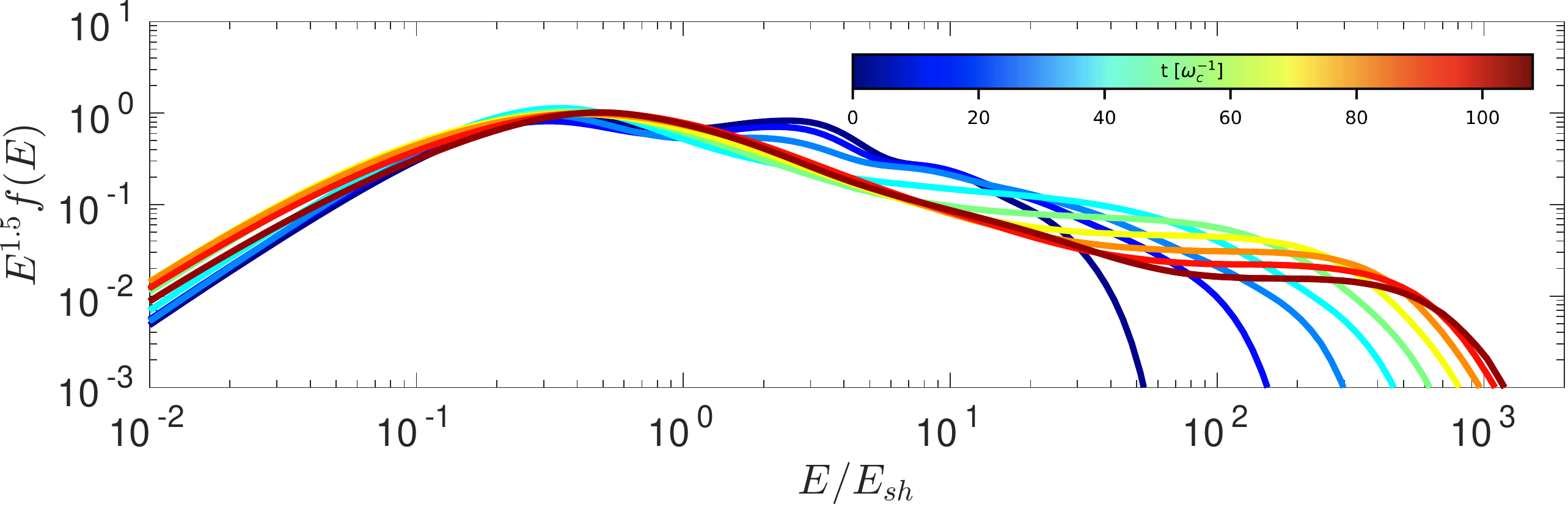"}
    \includegraphics[width=0.48\textwidth]{"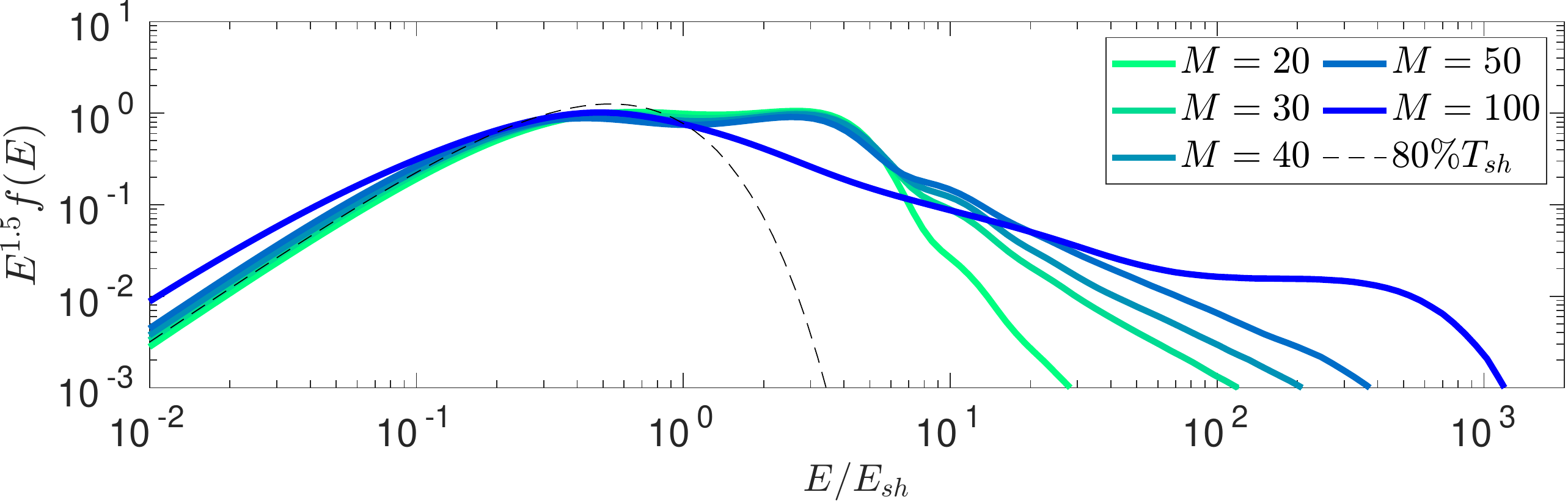"}
    \caption{Top panel: downstream ion spectrum  at $t=108 \omci$ for $M=100$ and different setups.
    A non-thermal tail above $10\esh$ is present only in 3D, while  1D and 2D show similar truncated spectra. 
    Middle panel: time evolution of the ion spectrum in 3D. 
    Bottom panel: spectrum dependence on $M$; the dashed line indicates a Maxwellian with temperature $\sim 80\%$ of the one expected for a purely gaseous shock.} 
    \label{Fig:Spectrum_example}
    \end{center}
\end{figure}

\emph{Magnetic field amplification.}
Fig.~\ref{Fig:B_field} shows the total magnetic field intensity $B_{\rm tot}$ for a shock with $M=100$ and $\thbn=80\deg$ in 2D$z$, 2D$y$, and a slice in the $x-y$ plane of the 3D setups;
the bottom panel shows the evolution of $B_{\rm tot}$ for the 3D case.
There is a striking difference between 2D$z$ and 2D$y$: 
while in the former case the field is simply compressed at the shock, in the latter case $B_{\rm tot}$ overshoots and is strongly amplified downstream  \cite{bohdan+21,kato+10,caprioli+14a};
at the latest stages of the 3D simulation the field amplification extends also upstream of the shock, a signature of a precursor produced by back-streaming particles.

The full characterization of the strongly amplified fields ($B_{\rm tot}/B_0\gtrsim 40$) observed for $M=100$ is beyond the goal of this Letter, though we notice that their morphology and scaling ($B_{\rm tot}/B_0\propto \sqrt{M}$) bear strong resemblance with those produced by the ion-Weibel instability \cite[see][and references therein]{kato+10,bohdan+21,matsumoto+15}.
Finally, we ascribe the difference between 2D$z$ and 2D$y$ to the fact that there is a finite baroclinic term, which in 2D can only have a $z$ component;
this hints to a contribution from turbulent dynamo processes, too.

\emph{Particle spectra and dependence on $M$.}
The top panel of Fig.~\ref{Fig:Spectrum_example} shows the post-shock ion spectrum at $t=108\omci$ for different setups: 
1D and 2D exhibit only a supra-thermal bump \citep{caprioli+14a}, while 3D shows a very extended power-law tail;
the dashed line corresponds to a Maxwellian with temperature $\sim 80\%$ of the one expected for a purely gaseous shock, which suggests that $\sim 20\%$ of the shock ram pressure is converted in energetic ions. 
Note that the post-shock 2D$y$ magnetic turbulence, while very similar to the 3D one, is not sufficient to grant injection into the acceleration process.
The evolution of the downstream ion spectrum in 3D is shown in the middle panel of Fig.~\ref{Fig:Spectrum_example}:
from an early supra-thermal bump, similar to the 1D/2D cases, a non-thermal tail develops very quickly and extends over more than three orders of magnitude in just $\sim 100\omci$.

Finally, the bottom panel of Fig.~\ref{Fig:Spectrum_example} shows the spectrum for different Mach numbers.
Each spectrum can be characterized as a power-law,  $f(E) \propto E^{-\alpha}$,  with an acceleration efficiency $\varepsilon$, defined as the fraction of the post-shock energy density in ions with energy above $10\esh$.
The scaling of $\alpha$ and $\varepsilon$ on $M$ are reported in Tab.~\ref{Table}. 
For $M$=100 the spectrum tends to $E^{-1.5}$ ($\propto p^{-4}$ for non-relativistic particles), the universal spectrum expected at strong shocks;
for lower values of $M$, non-thermal tails become steeper and less extended, almost vanishing for $M=20$ \citep[no tails were found for 3D simulations of $M=6$, see][]{caprioli+14a}.
The post-shock magnetic turbulence scales $\propto \sqrt{M}$, which increases the probability that ions return upstream and makes spectra harder.
Overall, the acceleration efficiency increases from a few percent for $M=20$ to $\varepsilon\gtrsim 20\%$ for $M\gtrsim 50$, a value comparable to the efficiency of quasi-parallel shocks \cite{caprioli+14a,caprioli+14b,haggerty+20}.

\begin{figure}[t!]
    \includegraphics[width=0.48\textwidth, clip=true,trim= 0 0 1 1]{"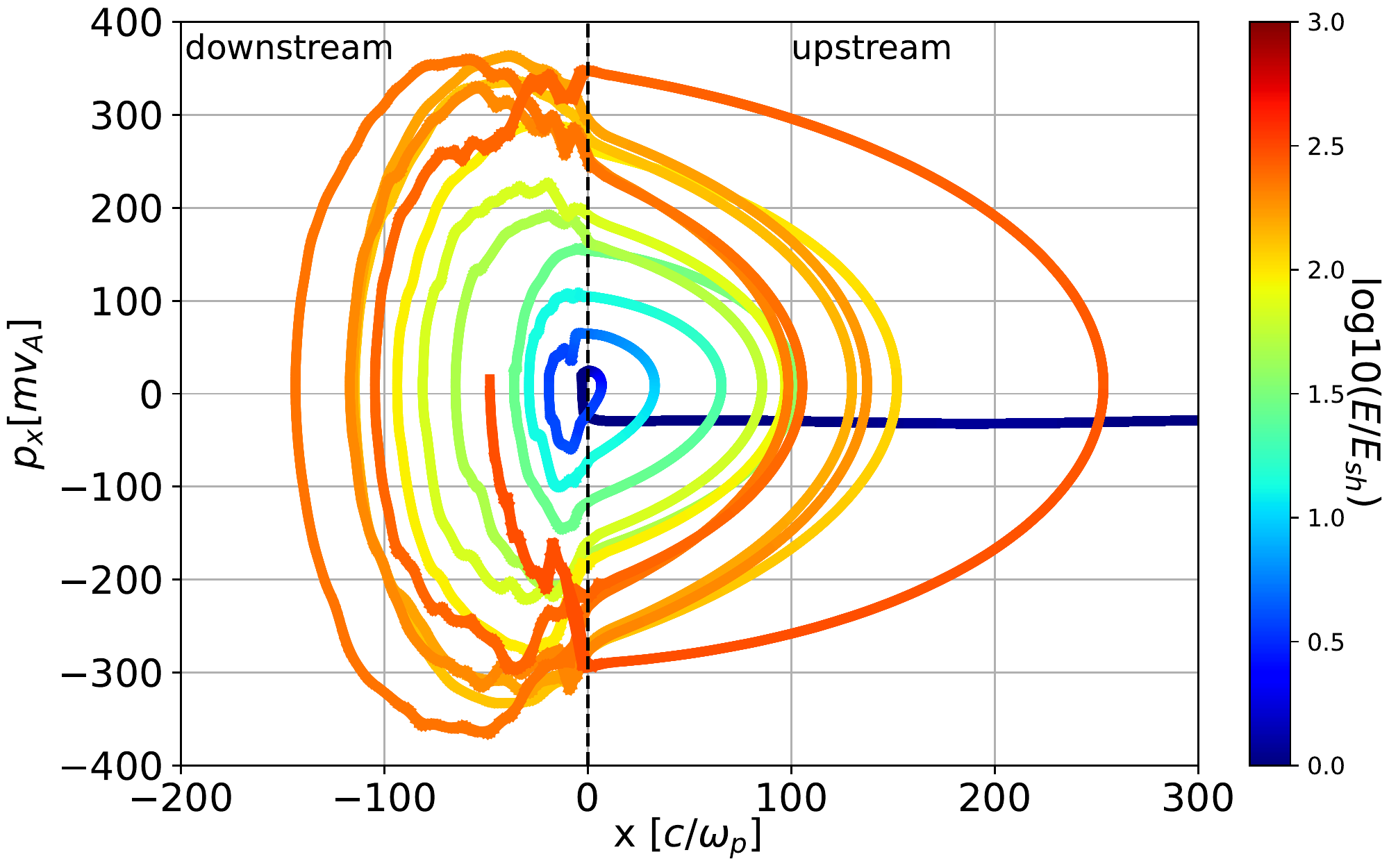"}
    \caption{Energy gain (color code) and trajectory in the $x-p_x$ plane of a representative ion undergoing SDA in run B.} 
    \label{Fig:Tracking_energy}
\end{figure}

\emph{Acceleration mechanism.}
Let us focus now on the mechanism(s) responsible for ion acceleration.
Fig.~\ref{Fig:Tracking_energy} shows the $x-p_x$ trajectory and energy gain of an ion tracked in run B. 
Particles gain energy through shock drift acceleration (SDA), tapping into the motional electric fields $\mathbf{E}\simeq - \mathbf{v}/c \times \mathbf{B}$ during their gyrations around the shock \cite{chen+75,decker+85,ball+01,yang+09}.

As pointed out in \cite{jokipii+93,jones+98,giacalone+00}, cross-field diffusion plays a crucial role in the return of ions from downstream, and is not properly captured if not in 3D.
\citet{jones+98} demonstrated that charged particles in an arbitrary electromagnetic field with at least one ignorable spatial coordinate remain forever tied to a $B$-field line.
Since in 2D field lines are effectively transverse ``sheets", ion diffusion along the shock normal is inhibited;
in 3D, instead, field lines can twist and intertwine, and ions can diffuse cross-field, which effectively prevents them from being rapidly swept downstream.
Tracking reveals that in 2D ions are advected downstream after a couple of gyrations, while in 3D they diffuse back several times, gaining energy at each SDA cycle. 
To some extent, this acceleration mechanism is similar to that proposed by \cite{kamijima+20}, who argued that the extreme case of a perpendicular shock where Bohm diffusion was realized downstream would be a rapid accelerator;
our self-consistent simulations show that the process may occur only for large $M$ and may be intrinsically limited when $\thbn<90\deg$.

\begin{figure}[t!]
\begin{center}
    \includegraphics[width=0.48\textwidth, clip=true,trim= 2 4 1 1]{"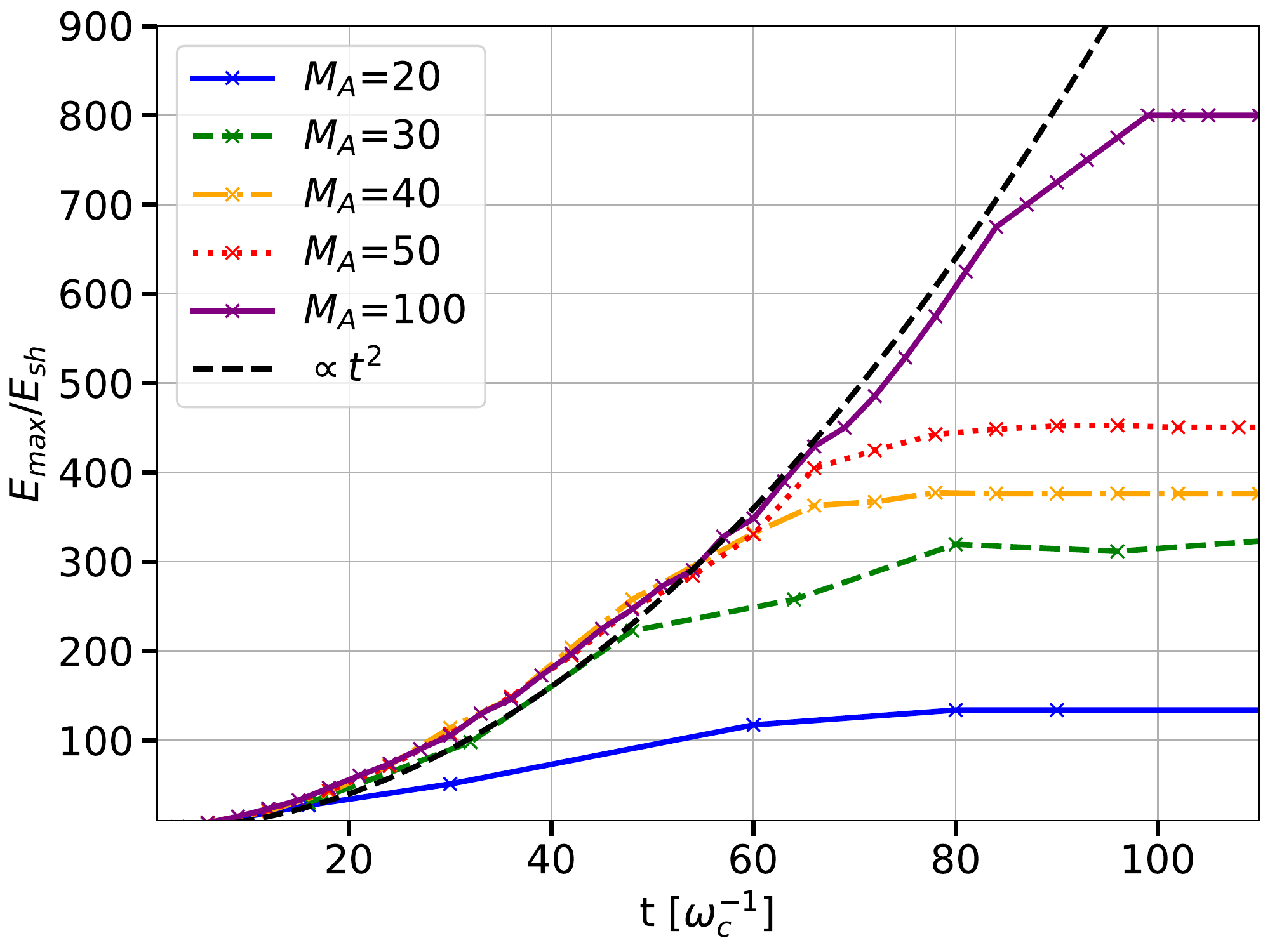"}
    \caption{Evolution of $\Em(t)$ for different $M$; after an initial increase $\propto t^2$ provided by SDA (black solid line), $\Em(t)$ eventually reaches an asymptotic value $\propto M$.} 
    \label{Fig:E_max}
    \end{center}
\end{figure}

\emph{A limit on the maximum achievable energy?}
Fig.~\ref{Fig:E_max} shows the evolution of the maximum ion energy $\Em$ for different $M$. 
After an initial growth $\propto t^2$, $\Em$ invariably saturates at an asymptotic value $\Em^*$, which scales $\propto M$ because ions can undergo more SDA cycles.

For non-relativistic ions, the momentum gain per SDA cycle is $\Delta p/p \propto \vsh/v$, where $v$ is the ion speed, which implies a constant $\Delta p \propto \vsh$.
Since the duration of each SDA cycle is $\Delta t\approx\omci$ (independent of $v$) and since the number of cycles $\propto t$, then $p_{\rm max} \propto \Delta p t/\omci \to \Em \propto t^2$, in good agreement with Fig.~\ref{Fig:E_max};
for relativistic particles, one would have $\Delta p\propto p $ and  $\Delta t\propto \omci p$, so $\Em\propto t$. 
This acceleration process is extremely fast  \cite[e.g.,][]{jokipii87,giacalone+94} and generally faster than DSA even when ions become relativistic, since in DSA the duration of a cycle is the diffusion time $\sim p/\vsh \omci$ \cite{drury83, caprioli+14c}. 
At later times and for larger values of $M$, we observe a brief transition from SDA to DSA, with some sufficiently-energetic ions returning to the shock via pitch-angle diffusion rather than via ordered gyration (note the deviation from the $\propto t^2$ curve before the plateau in Fig.~\ref{Fig:E_max});
eventually, ions escape toward upstream infinity and non-thermal tails stall at a critical energy $\Em^*\propto M$.
 
The crucial question is whether the streaming of escaping ions may trigger the Bell instability \cite{bell04} and self-support the acceleration to larger energies at later times, as at parallel shocks \cite{caprioli+14b}. 
From the bottom panel of Fig.~\ref{Fig:B_field} we infer that the precursor, where $B_{tot}/B_0 \gtrsim 1$ because of ion-driven instabilities, is  $\sim 1000 d_i$ wide, comparable to the upstream diffusion length of $\lesssim \Em$ ions. 
Further upstream, though,  $\delta B\ll B_0$  and there is hardly any fluctuations at scales resonant with $\Em^*$ ions, so their diffusion length quickly becomes (much) larger than the box, which makes it computationally prohibitive to follow the longer-term evolution of these systems with 3D hybrid simulations.

\section{Phenomenological implications}
The results described in this work have several applications in space/astrophysical shocks. 

\emph{Heliospheric shocks.}
MMS measurements at the Earth's bow shock show that quasi-parallel regions are generally more efficient than oblique ones at accelerating ions \cite{johlander+21,Lalti+22}; yet, efficiencies $\varepsilon\lesssim 10\%$ for $\thbn\gtrsim 70\deg$, compatible with our run at $M = 20$ (Tab.~\ref{Table}) and not with 2D ones which return $\varepsilon\lesssim 1\%$ \cite{karimabadi+14,caprioli+14a,johlander+21}.
Another application of the very fast acceleration that we observe may be in producing relativistic electrons at foreshock disturbances at the Earth's bow shock \cite{wilsoniii16b}.

\emph{Supernova remnants.}
Particularly interesting is the case of SN1006, which shows a bilateral symmetry defined by the direction of ${\bf B}_0$ \cite{rothenflug+04,bocchino+11}.
X/$\gamma$-ray emission comes from the quasi-parallel (polar caps) regions \cite{gamil+08,giuffrida+22,SN1006HESS}, implying the presence of multi-TeV electrons, while radio emission is more azimuthally symmetric \cite{rothenflug+04}, suggesting the presence of GeV electrons also in oblique regions.
While efficient ion DSA up to multi-TeV energies is consistent with the low polarization and strong synchrotron emission in the polar caps, electron acceleration in quasi-perpendicular regions is likely boot-strapped via SDA, as outlined here, and then can proceed up to GeV energies just because of the interstellar turbulence \cite{caprioli15p,blasi13}.
Whether \emph{multi-TeV} electrons should also be expected in quasi-perpendicular regions is an interesting question that hinges on the longer-term evolution of these systems. 

\emph{Radio SNe.}
Finally, we consider young extra-galactic SNe, whose radio emission  suggests that electrons are often accelerated with a spectral index $\alpha\simeq 3$ \cite[e.g.,][]{chevalier+06,margutti+21}. 
Such steep spectra at fast shocks $(\vsh\approx 10^4$  km s$^{-1}$) are hard to reconcile with standard DSA \cite[though see][]{bell+11,diesing+21}, but may be compatible with the results presented here, provided that $M\lesssim 50$ (see Tab.~\ref{Table}).
A SN shock propagating in the Parker spiral of the progenitor's wind may provide both relatively-small Alfv\'enic Mach numbers and quasi-perpendicular shock geometries \cite{chevalier+06}.
Furthermore, a flattening of the spectrum for higher shock velocity, that could imply higher $M$, has also been reported \cite{bjornsson+04,chevalier+06}.
Scaling the asymptotic $\Em^*$ illustrated in Fig.~\ref{Fig:E_max} by $\vsh\simeq 10^4$ km s$^{-1}$, one finds:
\begin{equation}
    \Em^* \simeq 0.22 {\rm GeV} \frac{M}{50}
    \left(\frac{\vsh}{10^4{\rm  km~s}^{-1}}\right)^2,
\end{equation}
i.e., radio-emitting particles may be produced very rapidly ($100\omci\lesssim$ one minute for $B_0\gtrsim 3$mG).

\section{Conclusions}
We used hybrid simulations to characterize ---for the first time in kinetic calculations without any \emph{ad-hoc} prescription for particle scattering and/or injection--- how ions can be very rapidly accelerated at non-relativistic, quasi-perpendicular shocks.
3D simulations are necessary to fully capture the amplification of the initial magnetic field (Fig.~\ref{Fig:B_field}) and the cross-field diffusion that allows ions not to be advected away downstream after a few shock crossings.
Acceleration starts via SDA, exhibiting a clear signature $\Em\propto t^2$, then briefly transitions to DSA (where $\Em\propto t$) before reaching a limit energy $\Em^*$, beyond which particles escape upstream. 

Acceleration efficiency and spectral slope strongly depend on the shock Mach number $M$ (Tab~.\ref{Table}): while for $M\lesssim 20$ efficiency is only a few percent and spectra are very steep, for $M\gtrsim 50$ efficiency can exceed 10--20$\%$ and spectra converge to the DSA ones, as flat as $p^{-4}$ in momentum;
also the level of magnetic field amplification and the maximum energy limit increase with $M$.

We have briefly outlined few applications of our results to space/astrophysical shocks; 
we defer to another work a detailed study of how injection and acceleration depend on $\thbn$ and the plasma $\beta$ parameter. 
The biggest questions that remain open are whether oblique/quasi-perpendicular shocks can efficiently drive plasma instabilities strong enough to self-sustain DSA up to energies significantly larger than $\Em^*$, and whether the same acceleration process is viable for electrons, too.
Both questions require different numerical approaches that are  capable of either capturing the longer-term evolution of the system or the physics of electron injection.

\begin{acknowledgments}
\section*{Acknowledgments} 
We warmly thank Lorenzo Sironi and Anatoly Spitkovsky for useful discussions.
Simulations were performed on computational resources provided by the University of Chicago Research Computing Center.
The work of L.O. is supported by the {\sc Departments of Excellence} Grant awarded by the Italian Ministry of Education, University and Research ({\sc MIUR}).
L.O. acknowledges support by the Research Grant {\sc TAsP} (Theoretical Astroparticle Physics) funded by the Istituto Nazionale di Fisica Nucleare. L.O. has been partially supported by ASI (Italian Space Agency) and CAIF (Cultural Association of Italians at Fermilab).
D.C. was partially supported by NASA through Grants No. 80NSSC20K1273 and 80NSSC18K1218 and NSF through Grants No. AST-1909778, PHY-2010240, and AST-2009326.
\end{acknowledgments}
\bibliography{Total}

\end{document}